\documentclass{aa}

\usepackage{graphics}

\begin{document}

%

\title {A catalogue of OB associations in IC\,1613}


\author
{
   J.~Borissova\inst{1}
   \and
   R.~Kurtev\inst{2}
    \and
   L.~Georgiev\inst{3}
     \and
   M.~Rosado\inst{3}
   }

\institute {Pontifica Universidad Catol\'{\i}ca de Chile, Facultad de F\'{\i}sica, 
Departamento de Astronom\'{\i}a y Astrof\'{\i}sica,
Av. Vicuna Mackenna 4860, 782-0436 Macul, Santiago, Chile.
Guest investigator of the UK Astronomy Data Centre
(jborisso@astro.puc.cl)
\and
Department of Astronomy, Sofia University and Isaac Newton Institute of Chile Bulgarian Branch, 
James Bourchier Ave. 5, BG\,--\,1164 Sofia,Bulgaria 
(kurtev@phys.uni-sofia.bg)
\and
Instituto de Astronom\'{\i}a, Universidad Nacional Aut\'onoma
   de M\'exico, M\'exico 
 (georgiev@astroscu.unam.mx,\  margarit@astroscu.unam.mx)
}

\offprints{J.~Borissova}

\date{Received 11 June 2001 / Accepted  15 September 2003}

\authorrunning {Borissova et al.}
\titlerunning {The catalogue of OB associations in IC\,1613}

\abstract{
We present a catalogue of OB associations in IC\,1613. 
Using an automatic and objective method (Battinelli's 1991 technique) 60 objects were found. 
The size distribution reveals a significant peak at about 60 parsecs if a distance
modulus of 24.27 mag is assumed. 
Spatial distributions of the detected associations and H II regions are strongly correlated.}

\maketitle


\section{Introduction}
 A stellar association  is "a single, 
unbound concentration of early-type luminous stars, embedded in a very young 
starforming region" (Kontizas et al. 1999).
The properties of the associations and the ionized gas clouds in which they 
are embedded allow tracing 
of the regions of most recent star formation in the galaxies.
IC\,1613 is a faint, irregular galaxy within the Local Group, which is
resolvable into stars stars from the ground.
The young stellar content of IC\,1613 was investigated first by Hodge (1978). 
On the photographic plates taken with the 5 m Hale telescope
he identified by eye estimates twenty OB associations. More recently, 
Freedman (1988) presented color-magnitude diagrams for 11 of them. 
Georgiev et al. (1999) presented the $UBV$ stellar photometry of 
the northeast sector of the galaxy. The properties of nine Hodge (1978) OB 
associations in this area are analyzed as well as seven new OB associations determined by cluster analysis.
Valdez-Gutierrez et al. (2000) presented  $\rm H\alpha$ and  $\rm[SII]$ observations using 
 the PUMA scanning Fabry-Perot interferometer. The kinematics of the ionized gas in the complex
 sample of nebulae was investigated. 
The ionized gas is distributed in classical HII regions and in a series of superbubbles.
They found that almost every superbubble in the NE region  has an interior association containing 
massive stars,
suggesting a physical link between them.  Rosado et al. (2001) continued the investigation of the 
influence of the
massive stars in the interstellar medium of IC\,1613 in the NW and southern  region of the galaxy.
 In the southern region they found that the 
superbubbles are probably formed by the winds of massive members of associations in spite of the 
close presence of a WO star. 
Lozinskaya et al. (2002) obtained spectra of the stars forming the chains and estimated their spectral
 types and luminosity classes. 
The stars were found to be at different evolutionary stages and six of them are identified as O stars. 
Lozinskaya et al. (2003) carried out detailed kinematical studies of the complex of multiple H~I
 and H~II shells
in  region of ongoing star formation in the dwarf irregular galaxy IC\,1613.

The purpose of the present paper, the fourth in our
IC\,1613 series, is to outline the new boundaries of the associations 
in IC\,1613 using observational material of uniform quality and an objective method 
of identification of stellar associations.

\section{Observations and data reduction}

A set  of  $UBV$ frames of IC\,1613 was obtained on the 2-m Richey-Chretien
telescope of the Bulgarian National Astronomical Observatory "Rozhen"  and the 2.1 m telescope 
of the San Pedro Martir  observatory in Mexico. The exposure times of the images are between 600 and 1800 sec. 
The field of view of both telescopes is  $5.6\arcmin \times\ 5.6\arcmin$. 
The seeing during these observations was $1 - 1.2 \arcsec$ with stable and  good photometric conditions. 
Three selected fields in the main body of the galaxy 
were observed. The Landolt (1992) standards were taken before and after all observations. 
The IRAF data reduction package was used to carry out the basic image reductions.
 The stellar photometry of the frames was performed with the point-spread function fitting routine 
ALLSTAR available in DAOPHOT (Stetson 1993). Complete details of the data reduction and analysis 
may be found in Georgiev et al. (1999).  
The zero-point errors of the transformations to the standard $UBV$ system are 0.04 mag. We 
transformed the $x, y$ coordinate systems of investigated fields to 
the one reference local system and then searched for stars in common  between  datasets. 
We check the residuals in magnitude and color for the stars in 
common. They do not show any systematic difference or trend so 
we can conclude that all datasets are homogeneous in magnitude within the errors. 
 The artificial star technique (Aparicio \& Gallart 1995) was used to investigate 
crowding effects. The completeness can seriously affect our analysis for magnitudes fainter than 21 mag. 
There are no significant differences in the completeness factors between investigated 
fields. Since our analysis concerns only bright OB stars, no completeness correction was made.
Field stars should not seriously affect the structure of the
color-magnitude diagram because IC\,1613 is situated far from the
Galactic plane and no correction for contamination by field stars was applied.
Fortunately IC\,1613 has very low reddening - 0.03-0.06
(Freedman 1988, Georgiev et al. 1999), which also dos not affects our results.

\section{Associations}

The automated search for OB associations was carried out using  
Battinelli's (1991) algorithm (also called the PLC technique).
The method assumes that "any given two stars in an ensemble of OB stars belong to 
the same association if and only if it is
possible to connect these two stars, by successively linking OB stars located 
between them, separated from each other by no more than a certain fixed distance 
parameter, or search radius, called $d_{\rm s}$". The distance parameter  $d_{\rm s}$ can 
be derived from the catalogue of early type stars using the function $f_{\rm p} (d)$.
This function describes the number of groups containing at least $p$ stars,
for any given value of the distance parameter $d$.
The function steeply rises until a maximum value is reached, followed by a gradual decrease
towards a limiting value of 1, which corresponds to the situation that the value of $d$ is so large 
that all stars can be assigned to one group. The optimum value of the distance parameter
 $d_s$ is defined as the 
value of $d$ corresponding to the maximum of the function $f_{\rm p}(d)$.  

In order to select the catalogue of blue stars we need to adopt brightness 
and color cut-offs. In IC\,1613 we selected as blue stars the stars that have
absolute magnitudes  $M_V < -2$
(assuming $(m-M)_{0} = 24.27\pm0.1$, Dolphin et al. 2000) and
dereddened colors $(B-V)_0 < 0.0$ (assuming $E(B-V) = 0.03$, Rienke and  Hodge 2001).
Recently Pietrzynski et al. (2001) used the same algorithm to investigate
associations in NGC\,300. They adopted Battinelli's 
(1991) method adjusting the two important
parameters in the Battinelli algorithm, described in the previous paragraph
 ---  distance parameter  $d_{\rm s}$ and
the minimum number of stars $p$ by numerical analysis.  
To derive the value of $d_{\rm s}$ Pietrzynski et al. (2001) applied
the PLC technique with several values for the distance parameter and several
values for the minimum number of stars, p. Their results show
that the position of the maximum of the function $f_{p}(d)$ 
depends slightly on p and is located in the relatively narrow interval 
(see their Fig.~2).  
The second parameter --- the  minimum number of stars p of the 
potential association was determined by Pietrzynski et al. (2001) by 
performing the statistical tests.
One hundred  random distributions of a number of stars equal to the number of
the sample of blue stars, and distributed over the same area, were created and
the PLC technique was applied to search for
potential groups. This experiment was repeated for different minimum numbers
of stars.  They found  that the spurious
detections are less than 10 \% if the six stars as a minimum 
population of OB stars of potential associations are chosen 
(see their Fig.~3).

Here, we follow  Pietrzynski et al. (2001) approach for IC\,1613. 
The value of $d_{s}$ was derived applying the method
with different values for the distance parameter. 
In our case the position of the maximum of
the function $f_{p}(d)$ is located in the interval  between 17   
and 23 pixels. The number of associations obtained with $d_{\rm s}$ = 17,18...22 and 23
pixels changes by  less than $5\%$.
Thus we confirm the Pietrzynski et al. (2001) 
result that the number of detected OB associations is insensitive to the adopted 
value of the search radius in this interval. 
We adopted $d_{\rm s} = 20 $ pixels as a search radius.
The next parameter that needs to be specified is the  minimum number of stars $p$ 
of the potential association. Again, using the statistical approach of Pietrzynski et al. (2001),
described in the previous paragraph
we derive four OB stars as a minimum number of stars in some clump in order to
have a "real" association.

With the above described parameters Battinelli's (1991) algorithm selects 58 associations 
with sizes between 30 and 130 pc 
(with accepted distance of 730 kpc, Dolphin et al 2000).

And finally to check our results and to cover the whole galaxy we retrieve
 the $UBV$ images of IC\,1613 
from the ING Archive, UK Astronomy Data Center, Cambridge. The images were 
obtained on 11 December 1998 with the 2.5m Isaac Newton Telescope and 
Wide Field Camera (WFC), which has a
 field of view of 11.4 x 22.8 arcmin. Unfortunately, they are not of 
high quality and not as deep as Rozhen images.
We reduced the images and transformed the instrumental magnitudes to the 
standard $UBV$ system
 using about 100 common stars between WFC and Rozhen photometry. Then we 
ran again Battinelli's (1991)
 algorithm in the same way as described above. 
On the WFC images we found 55 associations. 
Comparing with the associations obtained from Rohzen images there are 5 omitted 
associations ( G1, G2, G4, G8 and G26,  see Table ~1.)   
and 2 additional (G59 and G60) ones.
The WFC images are less deep than Rozhen data by approx. 1 mag in U and there are 
less than four OB stars in each of the omitted associations. Also, the boundaries of
 some associations in the fields in common 
 are slightly different. In general the average quantities of the common associations
 in the two lists however are little affected. 
The two additional associations are found in the areas not covered by the Rozhen frames.

The map of the association boundaries resulting from the automated search of IC\,1613 
is shown in Fig.~\ref{Fig01} and the properties of the associations are summarized 
in Table~1. Columns 1, 2 and 3 give the designation and the equatorial coordinates of 
the detected associations. 
The number of bright ($M_V < -2$) OB members in each of the associations, their sizes 
in parsecs and cross-identification with the catalogue of Hodge (1978) are given 
in the next columns.  The coordinates of the OB associations were determined as 
the mean coordinates 
of their separate member stars.

%
\begin{figure}[h]
 \caption{ Map of the associations in IC\,1613 outlined by  
Battinelli (1991) algorithm. 
}
 \label{Fig01}
\end{figure}

\begin{table*}\tabcolsep=1pt\small
\caption {Catalogue of OB associations in IC\,1613}
\begin{tabular} {lcccclcccccl}
\hline
Name& $\alpha_{2000}$ & $\delta_{2000}$ & size(pc)&  OB stars & Hodge & Name& $\alpha_{2000}$ & $\delta_{2000}$ & size(pc)&  OB stars& Hodge\\
\hline
G1&  01:04:53.92& +02:10:08.49&   33&     5 & - &   G32& 01:04:53.04& +02:06:11.00&     60&   7&H9\\  
G2&  01:04:54.11& +02:08:50.65&   46&     5 & - &   G33& 01:04:57.28& +02:05:15.55&     40&   10&H9\\ 
G3&  01:04:54.84& +02:10:30.98&   35&     5 & - &   G34& 01:04:59.08& +02:04:23.50&     60&   6&H9\\ 
G4&  01:04:57.74& +02:10:56.39&   30&     4 & - &   G35& 01:04:55.95& +02:05:44.97&     84&   18&H9\\  
G5&  01:04:58.04& +02:08:56.80&   27&     5 & -&    G36& 01:05:00.54& +02:05:17.01&     57&   7&H9\\  
G6&  01:04:58.54& +02:08:22.14&   43&     6 & - &   G37& 01:04:57.87& +02:04:46.41&     65&   7&H9\\
G7&  01:04:59.49& +02:09:17.81&   129&    36 & H10 &G38& 01:05:00.47& +02:05:06.94&     79&   13&H9\\ 
G8&  01:04:59.43& +02:08:40.54&   28&     4  & - &  G39& 01:04:53.61& +02:05:44.15&     72&   7&H8\\
G9&  01:05:00.26& +02:10:13.41&   33&     4 & - &   G40& 01:04:59.36& +02:05:52.03&     45&   6&H8\\ 
G10& 01:05:00.59& +02:10:42.84&   112&    36 & H11 &G41& 01:04:58.76& +02:05:24.97&     62&   8&H8\\ 
G11& 01:05:00.91& +02:09:41.73&   92&     14 &H13 & G42& 01:04:53.07& +02:04:56.66&     64&   4&-\\
G12& 01:05:01.21& +02:08:33.03&   40&     6 & H14&  G43& 01:04:49.11& +02:06:15.83&     44&   5&-\\
G13& 01:05:01.64& +02:07:31.51&   42&     6 & H14&  G44& 01:05:01.78& +02:06:19.08&     42&   6&H7\\
G14& 01:05:02.22& +02:08:44.37&   107&    27&H14 &  G45& 01:04:50.98& +02:04:29.19&     54&    6&-\\
G15& 01:05:02.22& +02:08:03.08&   112&    50& H15&  G46& 01:04:48.86& +02:06:47.12&     76&    8&H6\\
G16& 01:05:02.66& +02:09:25.29&   55&     8 & H13&  G47& 01:04:50.69& +02:07:01.43&     101&   10&H6\\
G17& 01:05:03.05& +02:09:38.14&   31&     7 & H13&  G48& 01:05:02.28& +02:06:01.92&     106&   9&H7\\
G18& 01:05:03.43& +02:08:35.60&   71&     11&H14 &  G49& 01:04:51.61& +02:06:23.82&     49&    4&-\\
G19& 01:05:04.17& +02:10:16.83&   51&     6 & -&    G50& 01:04:46.46& +02:07:28.28&     55&    4&H4\\
G20& 01:05:04.98& +02:11:55.88&   53&     7 & H16&  G51& 01:04:46.28& +02:08:59.07&     84&    6&H4\\
G21& 01:05:04.77& +02:09:31.83&   103&    23& H17,H13&G52&01:04:45.92& +02:06:28.20&     82&    6&H5\\
G22& 01:05:05.16& +02:11:14.26&   86&     10& -&     G53& 01:04:41.69& +02:09:16.98&    112&   10&-\\  
G23& 01:05:05.17& +02:08:50.48&   28&     5 & H17&   G54& 01:04:44.98& +02:09:24.31&    37&    5&-\\ 
G24& 01:05:06.07& +02:09:01.59&   45&     5 & - &    G55& 01:04:33.46& +02:08:21.15&    80&    6&H3\\
G25& 01:05:06.28& +02:09:30.08&   37&     11&H17 &   G56& 01:04:50.46& +02:08:14.68&    76&    5&H3\\ 
G26& 01:05:06.26& +02:06:52.73&   40&     5 & - &    G57& 01:04:43.40& +02:08:17.12&    76&    5&H2\\
G27& 01:05:06.71& +02:08:39.35&   72&     11&- &     G58& 01:04:46.84& +02:08:17.81&    90&    11&H1\\
G28& 01:05:07.59& +02:09:55.16&   111&    15&H17, H18&G59&01:05:10.41& +02:12:27.30&    60&    15&H20\\
G29& 01:05:10.05& +02:10:24.48&   68&     8 & H19&    G60&01:04:26.1 & +02:06:56.05 &    28&    4&-\\
G30& 01:05:13.68& +02:08:22.22&   41&     8 & - &             &     &             &   & \\
G31& 01:04:54.60& +02:06:04.61&   52&     6 &H9 &             &      &             &   &\\

\end{tabular}
\label{Tab01}
\end{table*}

Comparison with the stellar associations outlined by Hodge (1978) shows that most of them
coincide with our groups, but in general each Hodge association  divides into several smaller groups. 
The "new" associations have smaller sizes and look like bright cores within Hodge associations. 
Hodge (1986) describes the problem of the different sizes of the stellar associations in the Magellanic
clouds and other nearby galaxies as M\,31 and M\,33. He concluded that the mean size of the associations depends 
on the image scale and the distance to the galaxy. 
It is known also that young associations in the Galaxy, like the Trapezium, contain small compact subgroups
similar to those selected in IC\,1613 (Lada \& Lada 2003) while the older associations consist of randomly
 distributed stars and
they could not be divided into subgroups.
The Battinelli (1991) criterion obviously detects the youngest compact groups and this may explain 
the smaller mean size of the stellar groups identified in the present paper than those presented by Hodge (1978). 

\begin{figure}[h]
\resizebox{\hsize}{!}{\includegraphics{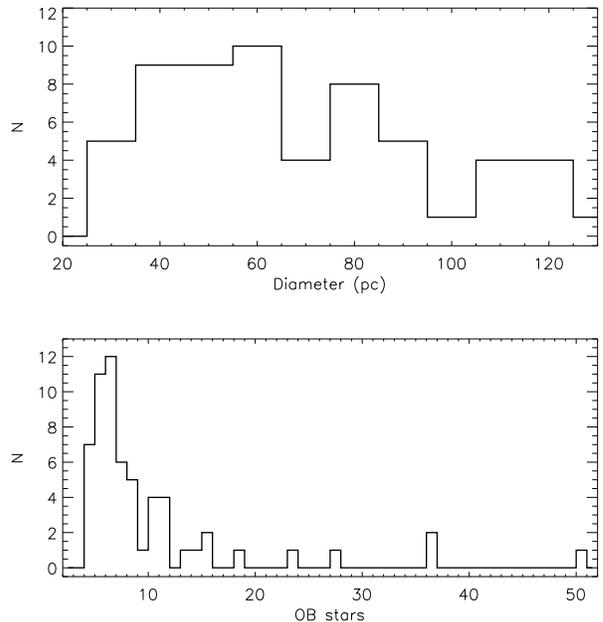}}
 \caption{ Top panel: Size distribution of associations in IC\,1613. 
Bottom panel: Histogram of the number of OB stars per association. 
}
\label{Fig02}
\end{figure}

In Fig.~\ref{Fig02} (top panel)  we show the size distribution of the detected associations. There is a peak located at 60 pc.
Comparing the IC\,1613 associations with the Magellanic Clouds, M\,31, M\,33 and NGC\,6822 
(Bresolin et al. 1998, Ivanov 1996)
associations we can see that the distribution of their sizes is similar with a peak between 40 and 80 pc. 
The  mean size of the associations in IC\,1613 is 63 pc and is in agreement with the mean sizes of associations in  
LMC - 60pc, SMC - 70 pc and M33 - 60 pc (Bresolin et al. 1998). All these measurements are performed 
using the same Battinelli (1991) method.
In order to check whether these similar sizes are caused by the identification algorithm we performed 
a numerical analysis.
Using DAOPHOT in IRAF we build the artificial images for our Field II. 
Field II contains 19 associations with a mean size of 64 pc and an average of 7 OB stars.
To conserve the completeness we chose to add 3 artificial associations with 7  "OB" stars.
We created these artificial associations with the PSF obtained
for Field II and added them randomly to the image. We then applied Battinelli's (1991) algorithm. 
We repeat this procedure many times. As a result we found that the size of an association does
not depend 
on how close the neighboring association is.
The same procedure was used to estimate the spurious detections 
of the associations due to random concentrations of blue stars. Here, we artificially added 21
"OB" stars ($10\%$ of the whole sample in  Field II) at random positions and created 20
images. Again, we apply Battinelli's (1991) algorithm and  estimated that between 3-5 OB 
associations in the whole sample could be randomly concentrated OB stars.

The histogram of the number of OB stars per association is shown in Fig.~\ref{Fig02}, bottom
panel. As can be seen most of the associations have fewer than twelve OB members, with a maximum
of the distribution at seven OB members. The most populous association contains 50 OB stars.

%

\section{Stellar associations and superbubbles}

The associations are not uniformly distributed within the galaxy and  they
are mostly concentrated in the NE and NW regions (Hodge 1978).  
On the other hand, the ionized gas in IC 1613 is concentrated in large diameter, expanding superbubbles 
(Valdez - Gutierrez et al. 2000) in the same regions. 
Comparison with the velocity map of the superbubbles (Figure~6 in Valdez-Gutierrez et al. 2000) 
shows that almost every superbubble in IC 1613 has an interior association. 
These associations are located within or at  the periphery of the superbubbles and  they
 are probably physically connected to the gas.
The map of the associations with superimposed boundaries of the superbubbles is shown in Fig.~\ref{Fig03}.

%
\begin{figure}[h]
 \caption{ Map of the associations in IC\,1613 outlined by  the
Battinelli (1991) algorithm (solid lines) with superimposed boundaries of superbubbles (dashed lines). 
}
 \label{Fig03}
\end{figure}

Some of the newly identified stellar
associations are found at the boundaries of the superbubbles and/or in the  interaction  zone between 
them.
This indicates new star formation at the edges of superbubbles.
Does this suggest, however, a scenario of self-induced star formation where the shocks that create the larger diameter 
superbubbles induce the formation of new stars compressing the ambient neutral or molecular gas? 
And are these more recently formed associations in turn creating new HII regions and bubbles
at the periphery of the original superbubble?
Lozinskaya et al. (2002) consider that the energy of the stellar wind from the associations is not
enough to form the ionized shells and found additional, intense sources of  stellar wind 
at their boundary --- early-type supergiants and giants. Lozinskaya et al. (2003) studied in detail
 the structure and kinematics of the neutral and ionized gas 
components in the complex of star formation in the NE region. They identified three extended 
(300\mbox{--}350~pc) neutral shells with which the
brightest ionized shells in the complex of star formation are associated. They also found evidence of the physical
interaction between the H~I and H~II shells in the
region of the chain of stars, early-type giants and supergiants, detected by Lozinskaya et al.~(2002). 
Taking into account the relative positions and ages of the H~I and H~II shells and OB~associations in the
complex, they suggested sequential or triggered star formation in the expanding neutral shells.
In addition to the three brightest and most prominent H~I shells, they found supergiant
arches and ring structures in the galaxy whose sizes are comparable to the gaseous-disk
thickness and  assumed this as a tracer of preceding starbursts in IC~1613.
Many OB associations however have no superbubbles. The gas  in these associations could be
evaporated or perhaps the ionization flux from the stars is not enough to ionize them.
In order to clarify these possibilities we are currently in the process of obtaining 
deeper photometry of the associations and performing spectral identification 
of the brightest blue stars. The results of such an analysis will be given in a
forthcoming paper.

\section{Summary}
We have presented the results of a search for OB associations in the dwarf irregular galaxy
galaxy IC\,1613. Application of the Battinelli (1991) method resulted in the
detection of 60 OB associations with sizes between 30 and 130 pc.
Numerical analysis indicates that between 3-5 OB associations in the
whole sample could be randomly concentrated OB stars. We detected the expected strong 
correlation between the spatial distributions of associations and H II regions in IC\,1613.

\begin{acknowledgements}
J.B is supported by FONDAP Center for Astrophysics grant number 150010003.
A part of this work was performed while J.B. was a visiting astronomer in 
the UNAM, Mexico.
The authors gratefully acknowledge the useful comments by an anonymous referee.
This research is partially based on data from the ING Archive.
L.G. acknowledges the financial support of the CONACyT No. 34422-E. 
J.B. thanks Marcio Catelan for his help. This research was supported in part by 
the Bulgarian National Science Foundation grant under contract No. F-812/1998 with 
the Bulgarian Ministry of Education and Sciences.

\end{acknowledgements}

\end{document}